# The Synthetic Mirror – Synthetic Data at the Age of Agentic AI

*Marcelle Momha*
*Harvard University, JFK School of Government*
marcelle_momha@hks.harvard.edu

**Abstract**

Synthetic data, which is artificially generated and intelligently mimicking or supplementing the real-world data, is increasingly used. The proliferation of AI agents and the adoption of synthetic data create a *synthetic mirror* that conceptualizes a representation and potential distortion of reality, thus generating trust and accountability deficits. This paper explores the implications for privacy and policymaking stemming from synthetic data generation, and the urgent need for new policy instruments and legal framework adaptation to ensure appropriate levels of trust and accountability for AI agents relying on synthetic data. Rather than creating entirely new policy or legal regimes, the most practical approach involves targeted amendments to existing frameworks, recognizing synthetic data as a distinct regulatory category with unique characteristics.

**Introduction**

For artificial intelligence (AI), data is the key driver of the explosion observed in recent years. The enormous and constantly growing volume of data driving this revolution does not come from real data in the usual sense of the word. Synthetic data, which is artificially generated and intelligently mimicking or supplementing the real-world data, is increasingly used. Indeed, a study by Gartner predicts that by 2030, synthetic data will dominate real world data in AI training [1]. Between 2026 and 2032, large language models (LLM) will be train on all available human text data on the Internet [2]. In 2022, the synthetic data generation market was estimated at US$288.5 million and will reach US$2,339.8 million by 2030 [3].

The rise of synthetic data is driven by the bottleneck observed in training large language models (LLM), which are immensely data- and resource-intensive. In addition, attempts to privacy preservation, data augmentation, and bias reduction provide additional motivations to address the vulnerability of real data. With the impending advent of Artificial General Intelligence (AGI), the proliferation of AI agents whose capabilities and applications will become increasingly widespread among the public and the adoption of synthetic data create a *synthetic mirror* that conceptualizes a representation and potential distortion of reality, thus generating trust and accountability deficits for agents created from and trained on artificially created data. How does the use of synthetic data fundamentally alter the challenges of establishing trust and accountability in AI agents? This paper explores the implications for privacy and policymaking stemming from synthetic data generation (SDG). It intends to show that existing AI and data governance frameworks are insufficient to address the unique challenges posed by the generation and use of synthetic data in agent training. There is an urgent need of new policy instruments and legal framework adaptation to ensure appropriate levels of trust and accountability for AI agents relying on synthetic data. In fact, contrary to the dichotomist narrative that typically presents the need for regulation and oversight as an obstacle to innovation, the real barrier to synthetic is public distrust when AI systems or agents could cause harm or violate expectations. Collaborative policymaking could provide the appropriate guidelines and principles to ensure the greater adoption of AI while ensuring the safety and protection of users.

**Methods for Generating Synthetic Data**

Although today most of the methods used to generate synthetic data use LLMs, it is important to mention the traditional statistical models also used.

1. **Generation of Synthetic Data using Statistical Models**
    a. **Parametric Modeling and Sampling**

This technique analyzes real data to determine the best fitting, known probability distribution (normal, exponential, chi-square). Once the parameters are estimated, synthetic data are generated by sampling and drawing from this fitted distribution. The goal is to simulate the characteristics of the original data in the sample values (with or without replacement), sometimes adding random noise to increase variability while preserving statistical properties [4]. Parametric bootstrap is a classic example [5]. Synthetic datasets are repeatedly sampled from the estimated model. Interpolation and extrapolation can also be used to generate new data points between or beyond existing data points.

   b. **Copula Models**

Copula models capture the dependence structure through correlations and other dependencies between multiple random variables in a dataset separately from their individual distributions. They can model complex dependencies between multiple variables [6]. After modeling these relationships and fitting a copula to the joint distribution, new data points are generated, preserving the statistical properties and correlation structure of the original data (Houssou et al., 2022; Syntheticus, 2025).

   c. **Monte Carlo Simulation**

This approach uses repeated random sampling from specified probability distributions to simulate possible outcomes and generate synthetic data sets [7]. It is widely used in domains like stress testing, risk analysis, and exploring the behavior of statistical estimators, as it allows for injecting a certain level of uncertainty into a real system.

### d. Stochastic Processes

Synthetic data can be generated by simulating stochastic processes (e.g., Markov chains, Poisson processes), which define probabilistic rules for the evolution of data over time or space [8].

## 2. Generation of Synthetic Data using Deep Learning & Artificial Intelligence

Advancements in deep learning these last years have led to the development of innovative method for anonymization and privacy protection through AI-generated synthetic data. Various AI, deep learning (DL), and machine learning (ML) techniques are used to generate synthetic data. These methods rely on advanced algorithms to learn patterns, distributions, and relationships within real-world datasets, enabling the creation of realistic artificial data.

### a. Generative Adversarial Networks (GANs)

GANs consist of two neural networks: a generator that creates realistic-looking synthetic data, and a discriminator that evaluates its authenticity. The generator creates synthetic samples, and the discriminator evaluates them [9]. Learning proceeds like a game, where the generator improves to deceive the discriminator. Both networks improve through competition, producing high-fidelity, realistic, and sharp synthetic data. GANs can produce highly realistic data, particularly in the areas of image and audio synthesis, video generation, dataset augmentation for training other ML models, and data anonymization. They can also capture complex and high-dimensional data distributions [10]. However, their training can be unstable and lead to mode collapse, with the generator producing only a limited variety of samples. They require careful architectural design and learning tricks to stabilize training [11].

### b. Variational Autoencoders (VAEs)

VAEs are unsupervised generative models with an encoder that compresses the data and a decoder that reconstructs it. By sampling from a latent space, VAEs can generate new data points that resemble the original dataset and exhibit similar characteristics. They are trained to optimize two objectives simultaneously: the correspondence between the decoded sample and the original input which encourages the learning of useful representations by the model, and the difference between the distribution learned by the encoder for a given input and a prior distribution [12].

VAEs are popular for image synthesis, anomaly detection, data augmentation, feature learning, and the generation of structured data such as molecules or text [13]. Their training is more stable than that of GANs [14]. They often produce better or sharper results than GANs, particularly for images. They can handle uncertain or missing data and generate diverse samples [15]. Their optimization objective involves approximation, which can limit generative quality.

### c. Transformers and Large Language Models (LLMs)

Transformers are a type of deep learning (DL) architecture that relies heavily on multi-head self-attention [16]. Originally developed for natural language processing (NLP), they excel at modeling sequential data and capturing broader dependencies. Models like the Generative Pre-trained Transformer (GPT) are specifically designed for content generation [17]. They are composed of stacked layers, each typically containing a multi-head self-attention module and a positional feed-forward network. Self-attention allows the model to weigh the importance of different parameters in the input sequence when processing a given element [18]. They are typically pre-trained on large datasets to predict the next element, like a token or a pixel, in a sequence based on all previous elements.

GPTs are the state-of-the-art solution for generating coherent and contextually relevant sequential data, especially text. They can be adapted to different data types and fine-tuned or optimized for specific tasks. However, their training, deployment and use are computationally expensive. They require enormous amounts of training data, which fuels the growing demand for synthetic data, and are likely to reproduce biases present in the training data. Furthermore, generation is inherently sequential and can be slow [19]. Models can sometimes *hallucinate* erroneous information or get stuck in repetitive loops.

### d. Simulation-based Generation

This approach generates synthetic data not by directly learning models from existing data, such as GANs, VAEs or Transformers, but by modeling the underlying process that creates the data using domain-specific models, physics-based simulation environments, mathematical models, agent-based systems, or other simulation frameworks [20]. It is commonly used in fields where collecting real data is expensive, dangerous, or impossible. The modeling phase defines the rules, equations, interactions, and environment that govern the required real-world data, drawing on domain expertise. The outputs, states, or events generated and recorded during the simulation constitute the synthetic dataset, which, in some cases, can become the industry's "ground truth" [21].

One of the strengths of this technique lies in its high fidelity if the underlying process is well understood and accurately modeled. Indeed, full control over the conditions and parameters of data generation during the simulation allows for the creation of specific, rare, or extreme scenarios where generating real-world data is dangerous or costly [22]. Weaknesses include the need for extensive domain expertise to build accurate simulations and collect high-quality data, which can lead to a "reality gap" when the simulation does not perfectly capture the nuances of the real world [23]. Running complex simulations can be computationally expensive, as experiment construction and validation are time-consuming.

### 3. Synthetic Data at the Age of Agentic AI

Currently, the world is preparing for a new way of thinking and experiencing human-machine interactions. Agentic AI and AGI are the next tech frontier. They will enable the deployment of autonomous decision-making systems capable of performing tasks and acting without human oversight or detailed instructions. This is the stage where AI reasons or *thinks* for itself, improving and adapting over time, learning from feedback and experience.

Given the growing data shortage, agentic AI systems will not only be built from primarily synthetic data but will also generate the data that will feed and train the LLMs [24]. AI agents could be assigned objectives related to data quality, diversity, coverage of specific scenarios, or usefulness for a downstream task. They could operate in a simulated environment and use their tasks or interactions to generate data. They could also learn or adapt their generative strategies based on feedback or experience to improve the quality or relevance of the synthetic data. This could create either a vicious or virtuous cycle, which effects and consequences remain to be determined. While the risks associated with LLM and agentic AI systems are the same as those mentioned above, their use to generate synthetic data could create new challenges.

### a. Autonomous & Self-Improving Data Generation AI Agents

Imagine a single, advanced AI agent whose primary goal is to generate synthetic data that meets certain quality criteria. This agent actively monitors the data it produces. Based on the evaluation, it uses learning algorithms such as reinforcement learning (RL), evolutionary strategies, online learning, or adaptive optimization techniques to autonomously adjust its internal generation process and improve results over time [25] [26]. Once the goals and initial configuration are defined, it operates with minimal human intervention.

One advantage of this method is that automation reduces the need for manual tuning and iterations in generative model development. It is easier to adapt to variations in the data distribution or changing requirements over time without complete retraining [27]. Exploration mechanisms within the learning loop could promote the discovery of new and useful patterns. By actively optimizing based on performance metrics, agents can produce more realistic or functionally relevant synthetic data. The downside is that the learning process can be unstable, potentially leading to model collapse or inconsistent data if not carefully managed. Optimization loops can be fragile and lead to instability, *catastrophic forgetting*, or suboptimal solutions if not carefully designed [28] [29]. Self-improving agents may inadvertently memorize details of real data, thus raising privacy concerns, or become highly specialized in certain tasks, to the detriment of broader utility [30] [31].

### b. Multi-Agent Systems (MAS) Simulation

This approach uses simulations powered by multiple interacting AI agents. Each agent has its own goals, behaviors – which can be rule-based or learned through RL – and interactions with other agents and the environment [32]. The synthetic data produced is a record of the agents' states, actions, and interactions over time that would reflect complex, real-world social, economic, or physical processes. Each simulation produces synthetic data that can be used to train or test models [33]. It is an extension or refinement of the simulation-based generation discussed above, emphasizing the agentic and interactive nature of the model.

An MAS simulation could uncover emergent behaviors and generate rich data from complex interaction patterns that are difficult to script manually. It would allow the creation of rare or extreme scenarios by adjusting environmental parameters or agent behaviors, and the replication of different populations with diverse agent types and behaviors. However, models trained on purely synthetic agent-based data can perform poorly in real-world situations if the simulation is not sufficiently realistic. This phenomenon is called *simulation-to-reality gap* [34]. Therefore, it could be extremely difficult to calibrate agents against real-world data and to validate that the simulation and agent behaviors accurately reflect reality. If agent rules or policies incorporate human biases, simulations can reproduce or amplify them. While capturing emergence is an asset, it can also be difficult to precisely predict or control the type of data that will be generated.

### c. Collaborative Cross-Domain Generative Agents

In this process, multiple specialized generative agents are involved, each expert in generating a specific type or modality of data – like text, images, audio, tabular data, time series. These agents collaborate, communicate, or coordinate their generation processes to produce coherent, multimodal synthetic datasets, where elements from different domains are logically connected [35]. Collaboration typically involves sharing partial results, constraints, or latent representations to produce coherent and integrated synthetic outputs. Multimodality enables the generation of rich datasets, useful for tasks such as visual and linguistic or audiovisual scenarios, with increased coherence and homogeneity within a synthetic sample. Each agent excels in its domain; complex real-world entities or events naturally involve multiple data types [36].

Collaboration can produce higher-quality and more coherent cross-domain data than a single monolithic model. Coordination complexity is a trade-off, as designing efficient communication and coordination protocols between agents is a major technical challenge. Agent failure or model drift can degrade overall data quality, as errors or artifacts generated by one agent can negatively impact the quality and consistency of other collaborative agents [37]. Achieving deep semantic and logical consistency across modalities is very difficult and often requires extensive world knowledge or constraints. Jointly training collaborative agents or an efficient orchestrator can be complex and data-intensive, creating a vicious cycle [38]. Ultimately, data exchanged between agents can inadvertently expose private information if the communication protocol is not secured or filtered [39].

**Foundations of Trust and Accountability in the Synthetic Mirror Governance**

The reliability of AI systems relies on multiple dimensions related to their deployment in real-world contexts. The increasing prevalence of synthetic data in AI development uniquely tests the limits of these dimensions.

### 1. Technical Reliability and Robustness

Reliability refers to the ability of an agentic system to consistently produce the expected and correct output under normal operating conditions. Robustness means that the system maintains its performance and functionality under various conditions, including unexpected or conflicting inputs and extreme cases [40].

Assessing the reliability of agents trained on generated data is one of the major challenges. When AI agents are trained on partially or fully synthetic data, reliable validation is difficult. A fundamental problem is validating behaviors learned from artificial data that are transferable to real-world scenarios. This creates what researchers call a *verification gap* due to the inability to fully guarantee that models learned from generated data reflect authentic relationships rather than artifacts of the generation process [41]. Traditional metrics assume the existence of concrete reference values or benchmarks, which may be lacking if models rarely consult real data or if *reality* is itself simulated. This could lead to *artificial confidence*, where AI agents overfit models from synthetic data, generating falsely high performance metrics that are not translatable to the real world. Furthermore, synthetic environments may omit edge cases or emergent behaviors, leaving agents ill-prepared for the complexities of the real world. They could also introduce systematic biases that may go unnoticed during the training phase but manifest as failure during the deployment. AI systems could learn to exploit patterns specific to synthetic data that do not exist in real-world environments.

### a. Lack of Comparison with Real-World Data

Another challenge with synthetic data lies in the lack of real-world data to allow for comparison. In traditional machine learning and deep learning development, models are validated against known real-world examples. With synthetic data, real-world datasets that can serve as validation benchmarks are often lacking. Synthetic data helps improve learning, but real-world data are necessary for true validation. The generative process itself can introduce unknown biases. It becomes difficult to distinguish legitimate models from generative artifacts. Furthermore, agents may learn shortcuts tailored to synthetic distributions, thereby ignoring cues from real-world data. Without access to the underlying data generation process, it becomes impossible to determine whether model behaviors represent real capabilities or simply the exploitation of synthetic models.

### b. Performance Gaps Between Synthetic and Real-World Environments

Models often exhibit degraded performance when moving from synthetic test environments to real-world applications. This gap is because the distribution of synthetic data can inevitably differ from that of the real world. Research regularly demonstrates performance disparities between systems evaluated on synthetic data and those deployed in real-world environments [42]. This gap can be particularly pronounced if the agents generating the data operate under simplified assumptions or if the simulation environment does not faithfully replicate reality. Thus, agents learn optimal strategies for the synthetic world, not necessarily for the real world. The gap between simulation and reality becomes particularly problematic in domains critical to safety and user protection, or when performance metrics on synthetic data tend to overestimate real-world capabilities. Even well-designed simulations or generative processes may fail to capture unpredictable real-world phenomena.

### c. Transparency and Explainability

Synthetic data adds another layer to the transparency and explainability challenge of synthetic mirrors. The generation process is often opaque. It can be difficult to explain precisely why the data appears the way it does or how a specific synthetic data point was generated [43]. If an agentic system trained on synthetic data fails in the real world, explaining this failure requires understanding the discrepancies between the synthetic data distribution and reality, which can be complex. On the other hand, the controllability of some synthetic data generation methods could potentially contribute to explainability by allowing researchers to generate specific data points or scenarios to systematically probe and understand some behaviors under controlled conditions.

Model designers, users, and policymakers need to understand how AI agents arrive at their decisions, especially when synthetic data or agent systems influence model behavior. Knowledge provenance becomes ambiguous when it comes from fabricated examples. Without clear pathways to explain decisions, systems trained on synthetic data face additional trustworthiness hurdles because stakeholders cannot confidently determine whether decisions stem from legitimate models or synthetic artifacts [44].

## 2. Ethical and Social Alignment

Ensuring that AI systems, whether agentic or not, operate in accordance with human values, ethical principles, legal requirements, and social norms is a priority and a topic of ongoing discussion. Among the most debated aspects are fairness, bias mitigation, and privacy protection.

### a. Fairness Through Fabrication and The Double-Edged Sword of Using Synthetic Data for Bias Mitigation

Synthetic data is sometimes touted as an innovative way to *rebalance* datasets that systematically lack the representativeness and diversity of real-world interactions, underrepresent or exclude important subgroups of a population [45]. The intention is to artificially increase the number of these underrepresented groups to create more comprehensive datasets [46]. This could potentially generate demographically balanced training data and mitigate biases stemming from skewed representation. Advanced techniques could generate data in which sensitive attributes such as ethnicity, gender, and health status are modified or decoupled from the synthetic data results, aiming to train less biased models. So, it would be possible to test counterfactual scenarios and verify the fairness of agentic systems.

While these techniques have several merits, oversimplification, excessive data manipulation, and poor modeling can introduce new biases or mask existing ones [47]. A generative model is not inherently neutral. It can introduce its own biases, conscious or unconscious, due to its architecture, training data, or optimization process if it encodes faulted assumptions. Oversimplified representations of marginalized groups can reinforce stereotypes. There is a risk of masking real inequalities if synthetic data are artificially balanced without accounting for actual systemic disparities. Downstream analyses risk ignoring real patterns of discrimination and providing a false sense of fairness without addressing structural issues.

### b. Privacy Requirements

Generating synthetic data that captures the statistical properties of real data without containing personally identifiable information (PII) is the holy grail and ultimate goal of synthetic mirrors. It would allow data to be shared for research or model training without exposing sensitive user information [48]. In reality, achieving true privacy is difficult.

Because AI systems memorize information, agentic models trained using synthetic data or generating the data themselves can inadvertently reveal sensitive trends or outliers if the generation process is not sufficiently randomized. Differential privacy guarantees are often weak or nonexistent when generating synthetic data, as they involve a trade-off between utility and data realism [49] [50]. This means that re-identification remains possible, especially with high-fidelity synthetic data. Hackers could reconstruct personal data by querying the model or analyzing the generated results.

### c. The Accountability Gap in complex AI systems

The increasing use of synthetic data in AI systems creates a critical flaw in our ability to assign responsibility and ensure adequate oversight when these systems fail or cause harm. The accountability gap stems from the complexity and intricacy of synthetic data pipelines, as well as the lack of established frameworks for monitoring, evaluating, and regulating artificially generated information.

The opacity of AI systems poses is another fundamental challenge to accountability [51]. In the event of actual harm, tracing the source of the problem becomes extremely difficult without adequate documentation of:

- The actors or agents responsible for creating the synthetic data.
- The source datasets that fed the generation process, including known limitations, potential biases, and any built-in fairness constraints or ethical guidelines.
- The specific models, algorithms, and parameters used, and the assumptions embedded in the training environment.
- Any post-processing, filtering, or fine-tuning applied after initial generation.

Synthetic data risks becoming a form of *data laundering* [52], where it is impossible to link problematic results to specific inputs or decisions. The outcome could be more concerning if synthetic data is used across organizational boundaries or integrated into complex AI systems where multiple synthetic datasets may interact in unpredictable ways [53]. Furthermore, the field lacks standards for reporting the use of synthetic mirrors (IEEE). Essential information regarding limitations and constraints often remains undocumented, and practices vary considerably across organizations and research groups.

The lack of standards and the ability of agent systems to autonomously adapt their own parameters transform synthetic data into an additional layer on top of AI algorithmic black boxes [54]. Without standards such as quality, reliability and fairness indicators, it is difficult to systematically compare datasets, assess their suitability for a specific task, or investigate failures. Documentation tends to be inconsistent and secondary, especially in rapid research and development environments. This limits the possibility of independent audits and verification, as technological devices structurally impede meaningful external oversight. Furthermore, audit methodologies quickly become obsolete as production technologies advance. Commercial interests can limit transparency and prevent scrutiny.

The dispersion of responsibilities among stakeholders in the generation, training, and deployment process complicates the attribution of legal liability when synthetic data causes or is linked to a failure or damage [55]. The supply and manufacturing chains of AI products and services are complex. Traditional liability models generally assume clear causal chains and identifiable responsible parties, assumptions that are virtually inapplicable to synthetic data pipelines. Legal frameworks struggle to adapt to the multi-stage technological processes inherent in AI [56]. The regulatory vacuum and lack of clarity reduce incentives to adopt rigorous, safe, and ethical practices, further undermining user trust.

**Existing Policy Landscape and Synthetic Mirror Policy Gaps**

To assess key national and state laws and policies related to AI and privacy, I conducted a basic text search on phrases such as "*artificially generated data*," "*synthetic data*," "*data generated by computer algorithms*," "*data generated by statistical models*," and "*AI-generated data*." Phrases "*AI-generated content,*" or "*synthetic content*" which are more common in these texts and refers to all sorts of AI-generated derivatives are intentionally avoided. I acknowledge that finding all global laws and policies mentioning these specific terms is a monumental task due to the rapid evolution of AI regulation, the large number of jurisdictions, language barriers, and the variable accessibility of legal documents. Moreover, some regulations could address this concept without using the exact phrases selected in the methodology. Therefore, this preliminary research focuses on the a few countries that were among that lead AI development or adoption.

The European Union (EU) AI Act is the only one to explicitly mention "*synthetic*" data in its corpus. Article 10 on data and data governance of high-risk AI systems imposes quality requirements for training, validation, and test data. Paragraph 5 stipulates that the "detection and correction of bias cannot be effectively ensured by processing other data, including synthetic or anonymized data" [57]. This suggests that legislators and policymakers are not anticipating the spread and impact of artificially generated data. The General Data Protection Regulation (GDPR), in Article 4(5), defines "anonymous data" as information not linked to an identified or identifiable natural person, which could, in some cases, apply to synthetic data but does not go further [58]. In the United States, no state or federal law, executive order, or policy framework uses or mentions

keywords related to synthetic data, not even the California Consumer Privacy Act (CCPA) which is one of the most comprehensive in the country [59].

In the United Kingdom, despite the complete absence of a national law on AI, the Office for National Statistics (ONS) has outlined certain broad considerations and requirements for the production and use of synthetic data for statistical research purposes and defined the roles and responsibilities of enforcement agencies [60]. In Singapore, the Personal Data Protection Commission (PDPC) has proposed a guide on the generation of synthetic data, which it describes as privacy-enhancing technology (PET) [61].

This near-total legal vacuum perfectly illustrates the need to put in place the necessary measures to oversee the generation and use of synthetic mirror data, in order to increase trust and adoption of innovations while protecting users. This is imperative, especially since the synthetic data market is booming.

**Proposed Policy Frameworks and Instruments: Polishing the Synthetic Mirror**

Filling the legal and policy gap in synthetic data production requires coordinated efforts among researchers, practitioners, market players and legislators. When crafting and implementing synthetic data policies, four main criteria can be used to assess their relevance: effectiveness, feasibility, enforceability, and adaptability. Any policy must ensure data quality, prevent harm, and facilitate the responsible use of synthetic data [62]. It must also promote the implementation of these standards and guidelines realistically in all organizations, regardless of their size and resources, and ensure that compliance costs are proportional to the expected benefits. It would also provide clear frameworks for monitoring violations and determining appropriate penalties, be flexible and adapt to different industries and AI applications [63], handle new methods and unforeseen challenges without becoming obsolete or hampering innovation. In this rapidly evolving field, two approaches are viable.

1. Data Privacy and AI Regulatory Updates

The first step is to update current data privacy and AI laws, signed and enacted, to explicitly encompass synthetic data, which is not limited to artificially generated multimedia or textual content [64]. With the advent of artificial general AI and the gradual introduction of AI agents into human-machine interactions, emerging AI regulations should specifically address systems trained exclusively, primarily or partially on synthetic data. Risk prioritization or tiering frameworks should consider the provenance of synthetic data to determine levels of control. Transparency obligations should include disclosure of the use of synthetic data and its potential impacts on system performance [65]. Testing and validation requirements must explicitly address synthetic-to-real world generalization gaps. Liability regimes must be updated to account for the diffusion of responsibilities in synthetic data pipelines.

## 2. Elaboration of Standards for Synthetic Mirrors

Standards development is a fundamental governance approach that can guide technology development by fostering confidence and trust in innovation. The Institute of Electrical and Electronics Engineers (IEEE) is currently working on creating technical standards defining minimum fidelity requirements that ensure reliability, robustness, and ethical and social harmonization [66]. Such initiatives could establish benchmarks for assessing the utility of synthetic data and standardize privacy-protective metrics and thresholds. These standards should mandate the documentation of the generation, training, and deployment processes. Information that may be tracked includes:

- Identification of the generating entity to the end user, statistical or generative models used,
- Actual source data used to train the generator, if applicable,
- Known limitations, biases, or privacy considerations of the source data, preprocessing steps applied,
- An explicit statement of assumptions built into the generator or simulation environment,
- Intended use cases for the synthetic dataset,
- Known limitations or areas where the data are not suitable,
- Results of quality, fidelity, utility, and bias assessments performed,
- A description of privacy-preserving techniques applied, such as differential privacy or others, and associated safeguards and limitations.
- A clear version control for datasets, as they are potentially updated or regenerated.

Beyond simple quantifiable metrics, comprehensive evaluation frameworks are also needed to assess synthetic data across multiple dimensions [67].

**Conclusion: Navigating the Future with Generated Realities**

Synthetic data generation offers different tradeoffs between fidelity and utility, along with other societal values like fairness and diversity. The choice of the method depends on the specific application and data characteristics. It depends on the nature of the data, project constraints and the required fidelity to real-world dynamics. Synthetic data, especially those generated by sophisticated LLMs and agentic AI methods, offers powerful opportunities to enrich datasets, explore scenarios, enhance privacy, and potentially mitigate bias. However, they pose significant challenges for establishing AI trustworthiness and accountability.

Trustworthiness assessment becomes more complex due to the lack of ground truth and the gap between synthetic and real-world data. Attempts at fairness can mask deeper issues or introduce new biases. The generation process adds layers of opacity, complicating transparency and explainability. Models can become brittle by overfitting to specific artifacts of the synthetic generation process. Therefore, the use of synthetic data requires extreme caution, rigorous validation against real data as much as possible, awareness of the limitations and potential biases of each method, techniques to bridge the gap with reality, and transparency about the data generation process itself. Synthetic mirrors bring considerable benefits, but their impact on reliability, transparency, explainability, bias mitigation and ethics must be constantly evaluated to reflect real-world values and foster trust and accountability throughout the value chain, from researchers to users. Rather than creating entirely new policy or legal regimes, the most practical approach to regulating or overseeing synthetic mirrors involves targeted amendments to existing frameworks, recognizing synthetic data as a distinct regulatory category with unique characteristics while leveraging established regulatory principles and enforcement mechanisms.

While this work makes a unique contribution to synthetic data research, it also presents limitations. A better understanding of agentic AI systems is needed, as this field is rapidly developing. Beyond the preliminary analysis conducted as part of this research, a broader and more in-depth text search using specially trained LLMs could uncover specific sectoral policies or regulations that may contain guidelines on synthetic data.